\documentclass[12pt]{article}
\usepackage{epsfig,amsbsy}
 
\newcommand{\mrm}[1]{\mathrm{#1}}

\newcommand{\MW}{M_{\mathrm W}}
\newcommand{\GW}{\Gamma_{\mathrm W}}

\newcommand{\lessim}{\raisebox{-0.8mm}%
{\hspace{1mm}$\stackrel{<}{\sim}$\hspace{1mm}}}
\newcommand{\alphas}{\alpha_{\mrm{s}}}
\newcommand{\alphaem}{\alpha_{\mrm{em}}}
 
\renewcommand{\b}{{\mathrm b}}
\renewcommand{\c}{{\mathrm c}}
\renewcommand{\d}{{\mathrm d}}
\newcommand{\e}{{\mathrm e}}

\newcommand{\p}{{\mathrm p}}
\newcommand{\q}{{\mathrm q}}
\newcommand{\s}{{\mathrm s}}

\newcommand{\B}{{\mathrm B}}

\newcommand{\K}{{\mathrm K}}

\newcommand{\W}{{\mathrm W}}
\newcommand{\Z}{{\mathrm Z}}
\newcommand{\bbar}{\overline{\mathrm b}}
\newcommand{\cbar}{\overline{\mathrm c}}

\newcommand{\pbar}{\overline{\mathrm p}}
\newcommand{\qbar}{\overline{\mathrm q}}

\newenvironment{Itemize}{\begin{list}{$\bullet$}%
{\setlength{\topsep}{0.2mm}\setlength{\partopsep}{0.2mm}%
\setlength{\itemsep}{0.2mm}\setlength{\parsep}{0.2mm}}}%
{\end{list}}
\newcounter{enumct}
\newenvironment{Enumerate}{\begin{list}{\arabic{enumct}.}%
{\usecounter{enumct}\setlength{\topsep}{0.2mm}%
\setlength{\partopsep}{0.2mm}\setlength{\itemsep}{0.2mm}%
\setlength{\parsep}{0.2mm}}}{\end{list}}
 
\newlength{\abstwidth}
\begin{document}
 
%set sloppy attitude to line breaks
\sloppy
 
\pagestyle{empty}
 
\begin{flushright}
CERN--TH/98--74\\
LU TP 98--8\\
hep-ph/9804202\\
March 1998\\
%change for version 2: add revision month
revised July 1998
\end{flushright}
 
\vspace{\fill}
 
\begin{center}
{\LARGE\bf Soft-Particle Spectra as a Probe}\\[3mm]
{\LARGE\bf of Interconnection Effects}\\[3mm]
{\LARGE\bf in Hadronic W$^{\boldsymbol{+}}$W$^{\boldsymbol{-}}$ %
Events}\\[10mm]
{\Large Valery A. Khoze\footnote{khoze@vxcern.cern.ch}}\\[3mm]
{\it INFN -- Laboratori Nazionali di Frascati,}\\[1mm]
{\it P.O. Box 13, I-00044 Frascati (Roma), Italy}\\[1mm]
{\it and}\\[1mm]
{\it TH Division, CERN,}\\[1mm]
{\it CH--1211 Geneva 23, Switzerland}\\[5mm]
{\large and} \\[5mm]
{\Large Torbj\"orn Sj\"ostrand\footnote{torbjorn@thep.lu.se}} \\[3mm]
{\it Department of Theoretical Physics,}\\[1mm]
{\it Lund University, Lund, Sweden}
\end{center}
 
\vspace{\fill}
 
\begin{center}
{\bf Abstract}\\[2ex]
\begin{minipage}{\abstwidth}
Cross-talk between the $\W^+$ and $\W^-$ sources of hadron production at 
LEP2 offers a hope to learn about basic properties of QCD, but at the same 
time threatens high-precision measurements of the $\W$ boson mass. Directly 
visible effects are not expected to be large, however. It is, therefore, 
important to develop methods to measure the level of interconnection 
in the data --- `connectometers'. In this article we study one particular 
set of observables, namely the reduced rate of low-momentum particle 
production caused by reconnection. In realistic scenarios the expected 
signals are at the border of visibility, but not quite beyond reach. Special 
studies on kaons and protons, or with other subsets of the data, could 
provide supporting evidence. We also show how LEP1 $\Z^0$ events could 
be used to define an alternative reference sample.  
\end{minipage}
\end{center}
 
\vspace{\fill}
 
\clearpage
\pagestyle{plain}
\setcounter{page}{1}

\section{Introduction}

The accurate determination of the $\W$-boson mass is one of the main 
objectives of LEP2. A statistical uncertainty on $\MW$ of about half
a per mille could eventually be achieved by reconstructing the $\W$ 
mass in hadronic $\W\W$ decays. However, the systematic uncertainties 
due to hadronic final-state interactions and QCD interferences 
between the $\W$ decay products may induce substantial ambiguities, 
for reviews see \cite{r1,r2,r3,r4}. Such an interconnection (cross-talk) 
between the $\W^+$ and $\W^-$ could occur because the
two $\W$ bosons decay at short distances of order $1/\GW \sim 0.1$~fm, 
and their decay products hadronize close to each other in space and time 
at the typical hadronic scale of $\sim 1$~fm.
The term `interconnection' is a generic name covering those
aspects of the final-state particle production that are not
dictated by the separate  decays of the $\W$ bosons, but can only be 
understood as a result of the joint action of the two.
  
The cross-talk between the $\W^{\pm}$ decay products undermines the 
traditional meaning of a $\W$ mass in the process
\begin{equation}
\e^+\e^- \to \W^+ \W^- \to \q_1 \qbar_2 \q_3 \qbar_4 ~,
\label{process} 
\end{equation}
called the (4q) mode.
It is not even in principle possible to subdivide the final-state 
hadrons into two groups, one of which corresponds to the 
$\W^+ \to \q_1 \qbar_2$ decay and the other to the 
$\W^- \to \q_3 \qbar_4$ decay: the identities of individual
$\W^{\pm}$ decay products are not well-defined any more.

The strong-interaction dynamics induces a variety of interconnection
effects between the hadronic decays of different $\W$'s, such as:
\begin{Enumerate}
\item
Quantum short-distance effects due to exchanges of perturbative gluons
between the two initial $\q\qbar$ systems.
\item
Final-state radiative gluon interferences on the stage of 
parton-shower development.  
\item
Long-distance effects in the parton-to-hadron transition phase 
caused by a large overlap between the products of the two decays
(non-perturbative rearrangement/reconnection).
\item
Bose--Einstein (BE) correlations between identical bosons
(in practice, pions).
\end{Enumerate}
  
As far as we are aware, the possibility of colour rearrangement in the 
process (\ref{process}) was first considered in \cite{r5}. 
The r\^ole of QCD interconnection in hadronic $\W\W$ decays in the 
framework of the $\W$ mass measurement was first
discussed in \cite{r6}.
This challenging topic has been quite intensively studied theoretically
since then, see e.g. \cite{r7,r8,r9,r10,r11,rHR}. 
BE correlations, in the W-mass context, were first discussed 
systematically in \cite{r12}, with further studies since 
\cite{r13,r14,rHR}.
However, in this paper we do not attempt to cover the consequences 
of BE effects.
The interest in QCD final-state interactions in the (4q) channel of
$\W^+\W^-$ decay has been strongly boosted by the prediction of the 
so-called colour-full scenario of \cite{r11}, that the shift in $\MW$ 
could be as large as several hundred MeV.

It is necessary to emphasize that there is no question of whether
interconnection between the $\W$'s exists or not; it is certainly there
even in the QED context.%
\footnote{The final-state QED interconnection induces a sizeable mass 
shift, $(\delta \MW)_{\mathrm{QED}} \sim O(\alphaem \pi \GW) \sim 50$~MeV,
in $\e^+\e^- \to 4$~fermions in the threshold region \cite{r15,r16}.
However, at energies above 170~GeV,
$(\delta \MW)_{\mathrm{QED}} \sim O(\alphaem \GW / \pi)$,
and cannot exceed a few MeV \cite{r16,r17}.}
Another well-known precedent is $\mathrm{J}/\psi$ production in 
$\B$ decay: the $\c\cbar \to \mathrm{J}/\psi$ transition requires 
a cross-talk between the two original colour singlets, 
$\cbar+\s$ and $\c + \mathrm{spectator}$.
The real challenge is to understand how large the ambiguities 
for various observables can be. Evidently, it is not only the $\W$ 
mass that can be affected by interconnection. Various event 
characteristics in hadronic $\W\W$ decay (such as the charged multiplicity 
or inclusive particle spectra) could, in principle, show effects even 
an order of magnitude bigger than that in $\MW$, see e.g. \cite{r6}.
On the other hand, in the inclusive cross section for process
(\ref{process}), the effects of the QCD (and QED) cross-talk are 
negligible \cite{r18}:
\begin{equation}
\frac{\Delta \sigma_{\W\W}^{\mathrm{intercon}}}{\sigma_{\W\W}} \sim
\left( \frac{C_F \alphas(\GW)}{\pi} \right)^2
\frac{1}{N_C^2 - 1} \frac{\GW}{\MW} ~,
\end{equation} 
where $C_F = (N_C^2-1)/2N_C$, $N_C=3$ being the number of colours.

A precise measurement of the $\e^+\e^- \to \W^+ \W^-$ threshold
cross section (see  e.g.\cite{r1} for details) would provide an
interconnection-free method for measuring $\MW$. Unfortunately, 
the combined total luminosity accumulated at LEP2 at a center-of-mass 
energy $\sqrt{s} = 161$~GeV (approximately 40~pb$^{-1}$ \cite{taylor}) 
is not sufficient to reach an interesting level of precision.
So the direct kinematic reconstruction of $\MW$ from the $\W$  
hadronic decays remains the only realistic method at current and 
future energies of LEP2.

The potential significance of the cross-talk phenomena for the $\W$ mass 
reconstruction at LEP2 obviously warrants a detailed 
understanding of the size of the corresponding ambiguities. Note also 
that QCD reconnection is of interest in its own right, since it may 
provide us with a prospective laboratory for probing hadronization 
dynamics in space and time.

The perturbative aspects of  QCD interconnection are, in principle,
well controllable. Since the corresponding $\W$ mass shift is expected 
to be well within the uncertainties of the hadronization models (and 
about on the same level as QED corrections) we only recall here an 
estimate of \cite{r6},
\begin{equation}
(\delta \MW)_{\mathrm{PT}} \sim
\left( \frac{C_F \alphas(\GW)}{\pi} \right)^2
\frac{1}{N_C^2 - 1} \GW ~,
\label{deltaMW}
\end{equation} 
which is of order of a few MeV. The perturbatively calculated 
mass-shift (as well as other observables) is colour suppressed, 
by two powers of $N_C$, which is typical for the gluon-mediated  
interaction between the two colour-singlet objects.

In the non-perturbative stage, which is our main concern in
this paper, the colour-suppression situation varies between
scenarios. Here factors like $1/N_C^2$ may present, 
as in the perturbative phase, but they are multiplied by 
model-dependent coefficients, which are functions of the space--time 
variables. These coefficients, in principle, could be 
anything, even much larger than unity. For instance, in the
models based on Lund string hadronization \cite{r19} the string itself 
consists of a multitude of coloured confinement gluons. Thus, if one 
gluon does not have the right colour to interact, then another nearby will.
In such scenarios the colour suppression factor could well be compensated.

Since the space--time separation between the $\W^+$ and $\W^-$ decay
vertices is typically of order $1/\GW$, only rather soft gluons (real
or virtual) with an energy $k \lessim \GW$ could feel the collective 
action of both the $\q_1 \qbar_2$ and $\q_3 \qbar_4$ antennae/dipole 
systems, and thus participate in the cross-talk.
This explains the origin of the last factor in eq.~(\ref{deltaMW}).
Non-perturbative reconnection, which is our main concern,
can occur wherever the hadronization regions of the two $\W$ bosons
overlap. As was first emphasized in \cite{r6}, the space--time 
picture of the evolution of the final state plays an 
essential r\^ole in understanding the size of the interconnection 
effects at the hadronic level. At the moment,
given a lack of deep understanding of non-perturbative QCD physics, 
the possible consequences of the hadronic cross-talk between the $\W$'s
can only be studied within the existing model-dependent
Monte Carlo schemes of hadronization. These have done a very good job
in describing a vast amount of information on hadronic $\Z^0$ decays,
so one may expect that (after appropriate modifications) they could 
provide a reasonable estimate for the magnitude of interconnection-induced 
effects.

The currently used algorithms for treating the non-perturbative cross-talk
\cite{r1,r2,r3} all assume a local interaction, i.e. the decay products of 
two $\W$'s are mainly affected in the regions of overlap. 
Reconnection-unrelated parameters are tuned to optimize the agreement 
with $\Z^0$ data. Some models (not studied in this paper) allow 
reconnection also among the partons of a single $\Z^0$, and then 
consistency requires reconnection to be included in the above-mentioned
tuning stage. The models  are based on different philosophies 
and include various assumptions, in particular, concerning the possible 
r\^ole of non-singlet hadronization.
Today  we are still very far from claiming that there is one `best'
recipe; rather each approach may manifest some particular aspects of 
the true physics. This does not mean that all models have to be put 
on an equal footing: some may be ``more equal than others''.

Some essential phenomenological aspects appear to be common for different
interconnection models:
\begin{Enumerate} 
\item
The cross-talk dampens comparatively slowly with center-of-mass
energy, $\sqrt{s}$, over the range that can be tested by LEP2.
\item
Interconnection effects tend to be strongly dependent on the event
topology, and could induce azimuthal anisotropies in the particle flow
distributions.
\item
The low-momentum final particles ($p \lessim 1$~GeV) are the main 
mediators in the hadronic cross-talk, and they are most affected by it.
\item 
Not far from the $\W\W$ threshold the invariant mass of an original
non-reconnected $\q\qbar$ system is larger than that for a  reconnected
one. Therefore, it is not surprising that most of the model predictions
show that the mean particle multiplicity in the (4q) mode,
$\langle N^\mathrm{(4q)} \rangle$, is lower than twice the mean 
multiplicity of a hadronically decaying $\W$ in the mixed 
hadronic--leptonic channel ((2q) mode), 
$\langle N^\mathrm{(2q)} \rangle$,% 
\footnote{Contrary to frequent statements in the literature, there is 
no logical need to expect BE effects to increase 
$\langle N^\mathrm{(4q)} \rangle$, thus potentially 
compensating the reconnection-based prediction (\ref{N42q})
(so-called BE conspiracy), see \cite{r14}.}
\begin{equation}
\frac{\langle N^\mathrm{(4q)} \rangle}{2\langle N^\mathrm{(2q)} \rangle}
 < 1 ~.
\label{N42q} 
\end{equation}
With increasing $\sqrt{s}$, the multiplicity in the purely hadronic 
final state may start to rise.%
\footnote{The latter effect is easy to understand considering an energy
far above threshold. Here both $\W$'s are strongly boosted away from each
other, and a reconnection between the two widely separated systems
leads to an increase in system masses, i.e. just opposite
to the threshold behaviour.}
However, at least within the models based on colour-confinement strings
discussed in \cite{r6}, the inequality (\ref{N42q}) remains valid in the 
whole range of LEP2 energies.
\item
All the models on the market (except of \cite{r11}) predict rather small
cross-talk effects \cite{r1,r2,r3,r4}. Thus, a conservative upper limit 
on the $\MW$ shift seems to be something like around 50~MeV. Changes 
in the standard global event characteristics are expected at the per cent 
level. In marked difference with all other approaches, the colour-full 
scenario of \cite{r11} allows much larger signals. Thus, the $\W$ mass 
and the relative multiplicity shifts are predicted to be around 400~MeV 
and 10\%, respectively. Note that the approach of \cite{r11} is the only 
available model which includes a non-singlet hadronization component.
The strong claims of \cite{r11} have made the whole subject of 
`connectometry' attractive for experimentalists.
\end{Enumerate}

The word connectometry is introduced here to cover various ways to
detect inter\-connection-induced effects by measuring 
characteristics of the $\W\W$ final state. The first experimental 
results on connectometry in the $\W^+\W^-$ events have already been 
reported (for a review see \cite{r20}), and new experimental 
information continues to pour out from LEP2, see e.g. \cite{r21}.
At the current level of statistics, there is no evidence for 
interconnection effects from the standard distributions in hadronic 
$\W\W$ events. This agrees with the mainstream of model predictions, 
which suggests rather small effects. However, it should be remembered 
that a $\W\W$ statistics larger by an order of magnitude is still 
to come.

An important point to bear in mind is that the values --- even the
signs --- of shifts in various observables can depend strongly on the 
hadronization scenario and on the choice of model parameters. Moreover, 
results may be strongly sensitive to the adopted experimental strategy 
(jet reconstruction method, event selection, etc.).

It would be extremely valuable to establish a model-independent
correlation between the shift in  $\MW$ and measurable quantities
in the final-state distributions. Unfortunately, so far studies do not 
suggest any convincing correlation of such a type. Moreover, it appears 
that the measurements of different observables in a real-life experiment  
may require different event selections.
So one has to proceed within the framework of a certain QCD Monte
Carlo model. This paper, as well as our previous ones \cite{r6}, is based 
on the models suggested by the Lund colour-confinement strings.
One of our main objectives here is to establish how the 
interconnection-induced signal could be seen through the eyes of 
such a simple inclusive observable as the low-momentum particle 
spectrum. We will also study a possible calibration procedure based 
on $\Z^0$ events.

\section{Soft-Particle Spectra as connectometers}

$\MW$ is not the best quantity for dealing with the interconnection
effects. Firstly, the $\W$ mass shift depends quite strongly on its 
definition (shift of the peak position of the mass distribution, or 
the average mass shift, or \ldots) as well as on the reconstruction 
strategy. It looks like quite a tough task to learn anything about 
the structure of the QCD vacuum from what is encoded in
$\delta \MW$ only. Secondly, the reconnection signature in
$\MW$ is not large --- it is only that we are set to measure this mass  
with the highest achievable accuracy that could make $\delta \MW$  
accessible. As  has been already mentioned, many event characteristics
may show relative effects, at least, an order of magnitude bigger.

The advantage of dealing with the standard global event shape 
characteristics for  connectometry purposes was understood 
already from the beginning \cite{r5,r6,r7}. It has also been found  
that an observable could become more sensitive to cross-talk if 
it is probed within some selected kinematical regions. For instance, 
such a quantity as the charged multiplicity may allow a better 
sensitivity to interconnection if measured in restricted ranges of 
thrust $T$ or in special rapidity intervals. However, there is 
always a price to pay, namely lower event statistics and a possible 
dependence of the signature on the jet reconstruction method
and experimental cuts.

There are several reasons to believe that the soft-particle yield
could provide a suitable candidate for the r\^ole of  connectometer:
\begin{Enumerate}
\item
It is quite a general consequence of the space--time picture of the
$\W\W \to 4\q$ process that the low-momentum particles ($p \lessim 1$~GeV)
experience most directly the cross-talk effects.
\item
The inclusive soft-momentum spectra of  particles $h$ in the isolated 
individual $\W \to 2\q$ decay, $\d n_{\W}^h / \d p$, can be rather 
well described, both within the QCD  Monte Carlo models and with 
analytical perturbative techniques \cite{r22}.
\item
The $\Z^0$ data provide an excellent experimental reference point, 
thanks to LEP1. When the $\Z^0$ results are used for calibration, the 
actual model dependence of the low-momentum spectra proves to be 
rather weak. Due to colour coherence in QCD cascades, the difference 
in the evolution  scales corresponding to the $\Z^0$ and the $\W$
could cause only small changes (on the per cent level) at low momenta,
see \cite{r23} and references therein. Effects due to the
difference in the primary quark flavour composition
also remain  on the per cent level for soft particles (see  e.g. \cite{r24} 
for a comparison between the $\Z^0 \to \b\bbar$ and $\Z^0 \to \q\qbar$ 
events). Such small corrections could readily be accounted for. 
\item
Inclusive soft-particle spectra are fairly independent of the jet
reconstruction strategy and event selection.
\end{Enumerate}

To implement colour reconnection we follow below the ideas and 
techniques that were developed in \cite{r6}.
Two main scenarios are called `Type I' and `Type II' in analogy with 
the two types of superconducting vortices which could correspond 
to colour strings. In the `Type I' models the reconnection probability 
depends on the integrated space-time overlap of the extended strings 
(elongated bags) formed in the two $\W$ decays. In the `Type II', the 
reconnection occurs at the crossing of two string cores 
(vortex lines). The  `Type II$'$' is a variant of the latter, with the 
requirement that a  reconnection is allowed only if
it leads to a reduction of the string length. We also present 
results corresponding to a simplified implementation (by us, using 
{\sc Jetset} \cite{r27} rather than {\sc Ariadne} \cite{ariadne} showers)
of the `GH' model \cite{r7}, where the reconnection is selected solely 
based on the criterion of a reduced string length. 
Numbers are also shown for two toy models, the `instantaneous'
and `intermediate' scenarios of \cite{r6}. In the former (which is 
equivalent to that in \cite{r5}) the two reconnected systems 
$\q_1\qbar_4$ and $\q_3\qbar_2$ are immediately formed and then 
subsequently shower and fragment  independently of each other.
In the latter, a  reconnection occurs between the shower and 
fragmentation stages. One has to bear in mind that the last two 
`optimistic' (from the connectometry point of view) toy approaches are 
oversimplified  extremes  and are not supposed to correspond to the
true nature. These scenarios may be useful for reference purposes, 
but their experimental rejection can at most be considered as 
warmup exercises for the real task.

Within these models there are some 
general qualitative predictions for the soft-particle spectra in the 
$\W\W \to 4\q$ events, $\d n_{4\q}^h / \d p$, in the LEP2 energy range.
\begin{Enumerate}
\item
Depopulation of the low-momentum hadrons, relative to the 
no-reconnection scenario, due to the Lorentz boosts of the alternative 
$\q_1\qbar_4$ and $\q_3\qbar_2$ dipoles/antennae.
\item
As in the case of the well-known standard string effect \cite{r19,r25},
such a depopulation should become more pronounced for heavier
hadrons (K, p, \ldots).
An unseparated sample of kaons and protons would also be less affected
by the BE phenomenon.
\item
A gradual reduction of the cross-talk with center-of-mass energy, since 
the two outgoing $\W$ hadronic systems are more and more boosted apart.
\end{Enumerate}

It is worthwhile to emphasize that a final-state observable  may
well have different sensitivity levels to different interconnection 
effects. Thus, connectometry based on the soft-particle yield is 
supposed to serve mostly as a hearing-aid for the long-distance 
hadronic cross-talk. The readers are reminded that the very idea of 
such studies has emerged from the string/dipole fragmentation picture.  
By contrast, it cannot be ruled out that BE effects will  not be 
seen in inclusive  spectra, and only emerge in particle correlations.
Furthermore, low-momentum spectra do not look like a proper diagnostic
tool to probe virtual gluon (photon) interference phenomena. As an 
instructive example, consider QED Coulomb effects on the $\MW$ 
reconstruction \cite{r15,r16}. While the $\W$ line shape is sizeably 
affected in the threshold region (at the 50~MeV level), it is quite 
unlikely one would get any significant distortions in the inclusive 
final hadron spectra.

\section{Numerical Results}

Armed with the algorithms of \cite{r6,r7} we can now proceed to 
calculate the effects of hadronic cross-talk on the low-momentum 
distributions.
Note that the momentum spectra in the following are considered 
as totally inclusive quantities. So all hadronic $\W\W$ events are 
included, without any cuts. It is up to experimentalists to select 
cuts which reduce the background and to correct for that.

Results have been obtained with {\sc Pythia} version 6.1 \cite{r27}.
All numbers are based on samples of 400,000 events per model.
The quoted number of decimals at times is larger than
warranted by statistical accuracy, in order to simplify comparisons.
Systematic errors, of course, by far exceed statistical ones, but
generally divide out in comparisons. (For instance, those coming 
from different tunes of fragmentation parameters to the LEP1 data.)

\begin{figure}[p]
\begin{center}
%change for version 2: corrected curve for intermediate scenario
\mbox{\epsfig{file=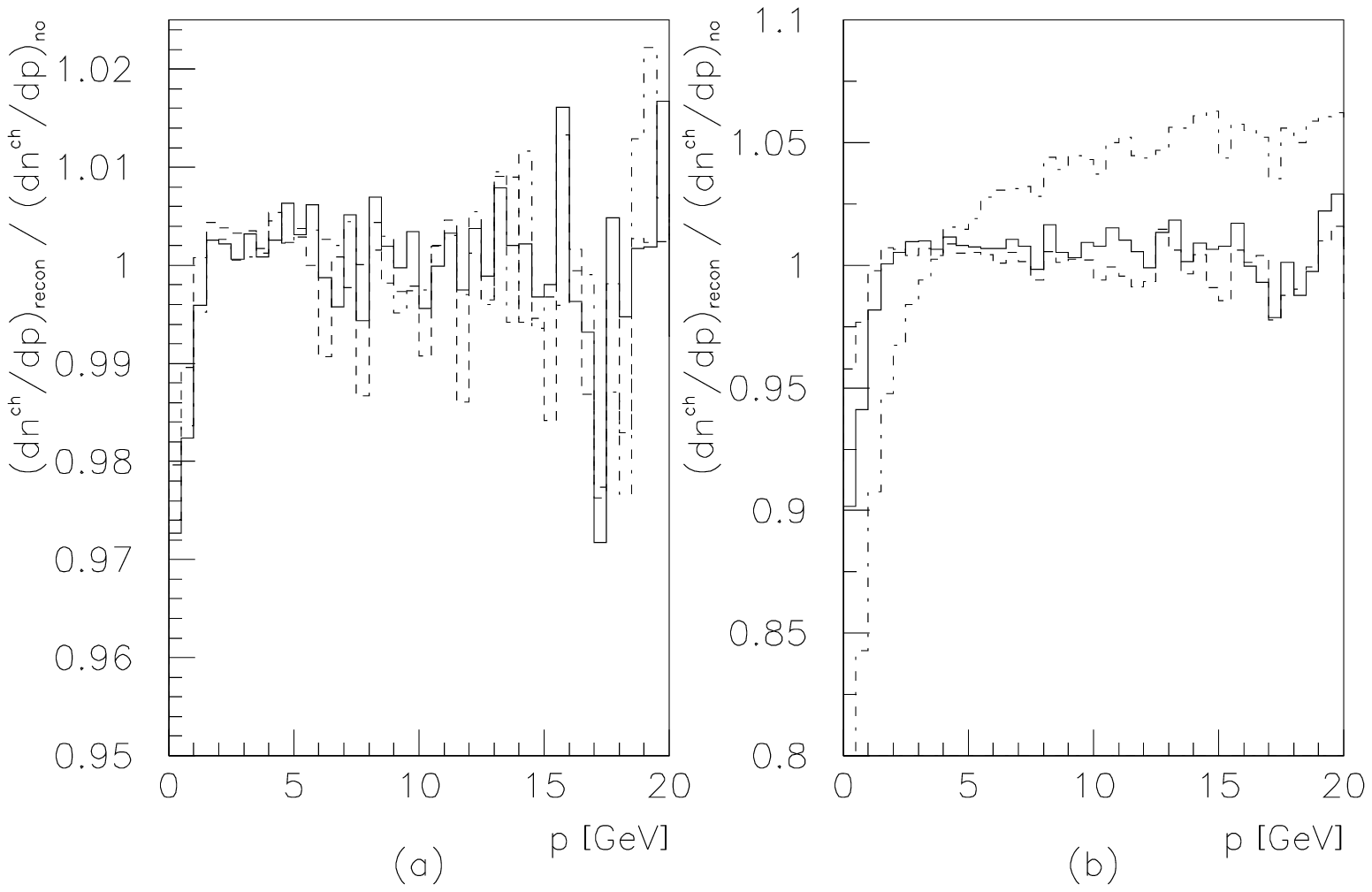}}
\end{center}
\vspace{-10mm}
\caption[dummy]{Ratio of reconnection to no-reconnection momentum
spectra $\d n^{\mathrm{ch}} / \d p$ for all charged particles with
$p < 20$~GeV. 
\textit{a)} Model I full, II dashed and II$'$ dashed-dotted.
\textit{b)} `GH' full, intermediate dashed and instantaneous dashed-dotted.
Note difference in vertical scales. Energy is 172~GeV; no ISR.}
\label{fig1}
\begin{center}
%change for version 2: corrected curve for intermediate scenario
\mbox{\epsfig{file=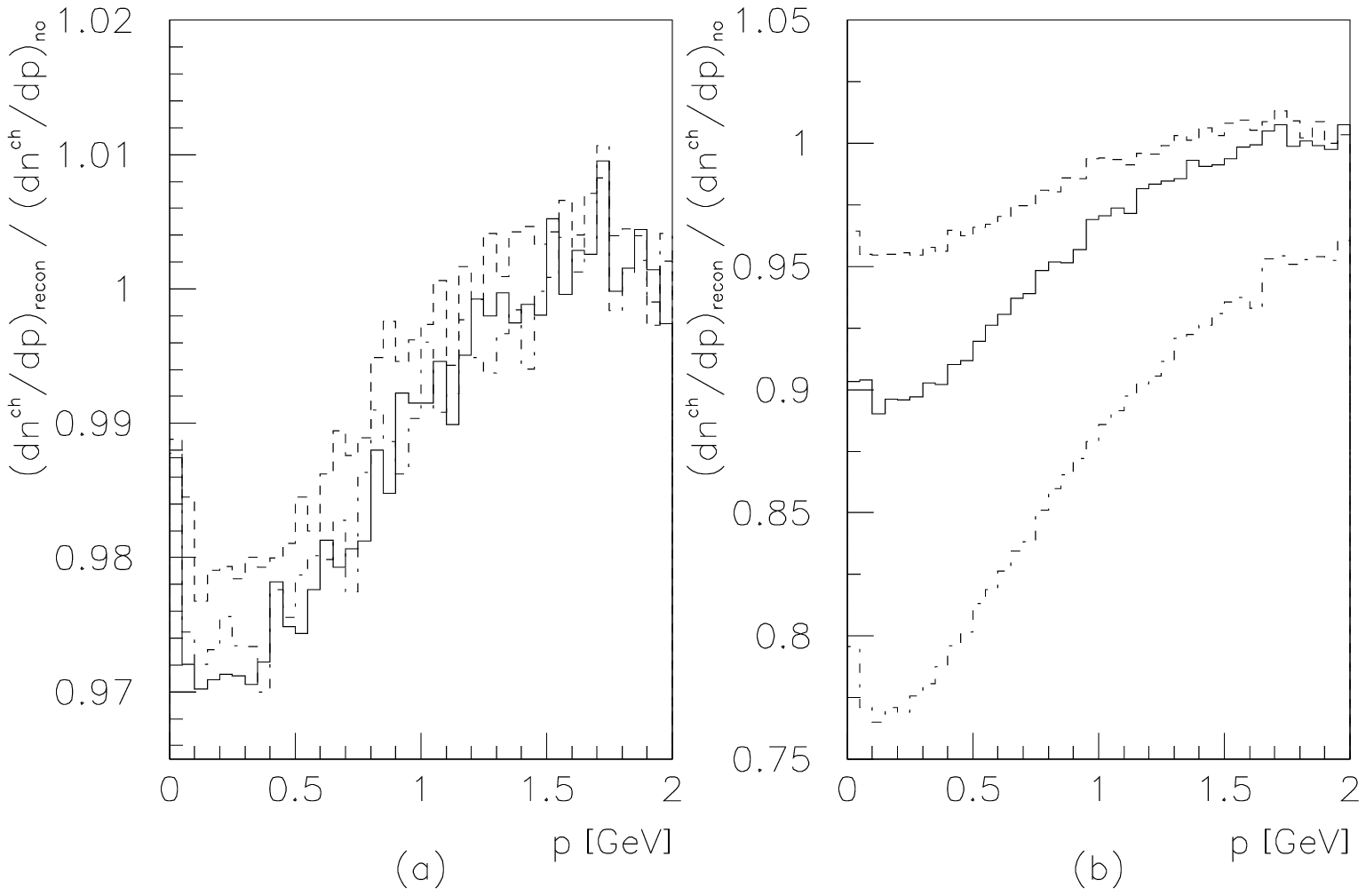}}
\end{center}
\vspace{-10mm}
\caption[dummy]{Ratio of charged momentum spectra as in 
Fig.~\protect\ref{fig1}, for the low-momentum region $p < 2$~GeV.}
\label{fig2}
\end{figure}

To illustrate the size of the expected effects, consider first the
energy  $\sqrt{s} = 172$~GeV.
The ratio of the momentum spectrum of charged particles with 
reconnection to that without it (a hypothetical no-reconnection 
case which is equivalent to the mixed leptonic-hadronic (2q) channel) 
is shown in Fig.~\ref{fig1} for various scenarios.
The nominal input $\MW$ is 80.33~GeV, and QED initial-state 
radiation (ISR) is not included. Since the main changes occur in the 
low-momentum range, we plot in Fig.~\ref{fig2} the same distributions 
as in Fig.~\ref{fig1} but only for momenta up to 2~GeV.

As expected, a suppression of the ratio at low momenta is quite 
visible. Note that the vertical scales are quite different in the
(a) and (b) frames. 
We have to warn the reader that the `GH' model contains the 
reconnection probability as a free parameter, here set to unity.
If one assumes reconnection at the 30\% level, as is the order 
obtained in the II and II$'$ scenarios and is assumed in the I one,
the 0.9 level of `GH' at very low values of $p$ would change to 0.97, 
which is comparable with the above three models. It is thus only the
%change for version 2: remove intermediate scenario
%unrealistic intermediate and instantaneous scenarios that deviate
unrealistic instantaneous scenario that deviates
from the rest, by giving significantly larger suppressions. 

The `shape universality' among the realistic models could be
related to all of them being based on the same string fragmentation
framework, but we would like to believe that it may be more general
than that. (Obviously a point well worth studying further.) If so,
we here have a good gauge for the reconnection probability. However, 
remember that models with the same probability may well give different
$\W$ mass shifts, so we are still not gauging the key quantity. 

Note that the phase space for particle production behaves like
$p^2 \d p$ at small $p$. The region $p < 0.2$~GeV therefore contains
few particles and is sensitive to statistical fluctuations.
The large spikes seen in the first bin or two of Fig.~\ref{fig2}
should therefore not be overinterpreted. In particular, the
same no-reconnection event sample is used as denominator for all the
ratios, so a downwards fluctuation here could be the common cause
for all the spikes.

To further quantify the expectations we give in Tab.~\ref{tab1} the 
results for the mean charged multiplicity
$\langle N_{\mathrm{ch}}^{\mathrm{(4q)}} \rangle$
together with the  charged multiplicity in the momentum range 
$p<1$~GeV, $\langle N_{\mathrm{ch}}^{\mathrm{(4q)}} \rangle%
|_{p < 1~\mathrm{GeV}}$.
The multiplicity difference between models is almost entirely coming 
%change for version 2: remove intermediate scenario
%from $p < 1$~GeV, except in the intermediate and instantaneous scenarios,
from $p < 1$~GeV, except in the instantaneous scenario,
where reconnection effects extend further out in momentum.
Some increase of multiplicity outside of the centrally depleted region
is a direct consequence of energy conservation, however. 

\begin{table}[tbp]
\begin{center}
\begin{tabular}{|l|c|c|c|c|}
\hline 
   & \multicolumn{2}{c|}{$\sqrt{s} = 172$~GeV} &
\multicolumn{2}{c|}{$\sqrt{s} = 195$~GeV} \\
\hline
model & $\langle N_{\mathrm{ch}}^{\mathrm{(4q)}} \rangle$ &
$\langle N_{\mathrm{ch}}^{\mathrm{(4q)}} \rangle |_{p < 1~\mathrm{GeV}}$ &
$\langle N_{\mathrm{ch}}^{\mathrm{(4q)}} \rangle$ &
$\langle N_{\mathrm{ch}}^{\mathrm{(4q)}} \rangle |_{p < 1~\mathrm{GeV}}$ \\
\hline
no-reconnection & 38.264 &  15.141 &  38.385 &  14.175  \\ \hline
I               & 37.941 &  14.802 &  38.230 &  14.025  \\
II              & 38.070 &  14.909 &  38.284 &  14.063  \\
II$'$           & 37.958 &  14.822 &  38.223 &  14.009  \\ \hline
`GH'            & 37.108 &  13.955 &  37.685 &  13.502  \\
%change for version 2: corrected numbers for intermediate scenario
%Intermediate    & 35.259 &  12.631 &  35.597 &  12.227  \\
Intermediate    & 37.844 &  14.648 &  38.684 &  14.519  \\
Instantaneous   & 34.929 &  12.303 &  36.770 &  12.749  \\ 
\hline
\end{tabular}
\end{center}
\caption%
{Charged multiplicity, in total and for $p<1$~GeV, at 172 and 195~GeV,
respectively. No ISR.
\label{tab1}}
\end{table}

Note that the averaged measured value of 
$\langle N_{\mathrm{ch}}^{\mathrm{(4q)}} \rangle$ at 
$\sqrt{s} = 172$~GeV \cite{r20},
\begin{equation}
\langle N_{\mathrm{ch}}^{\mathrm{(4q)}} \rangle_{\mathrm{exp}} =
38.94 \pm 1.07 ~, 
\end{equation}
is in a good agreement with the no-reconnection prediction.

\begin{figure}[p]
\begin{center}
%change for version 2: corrected curve for intermediate scenario
\mbox{\epsfig{file=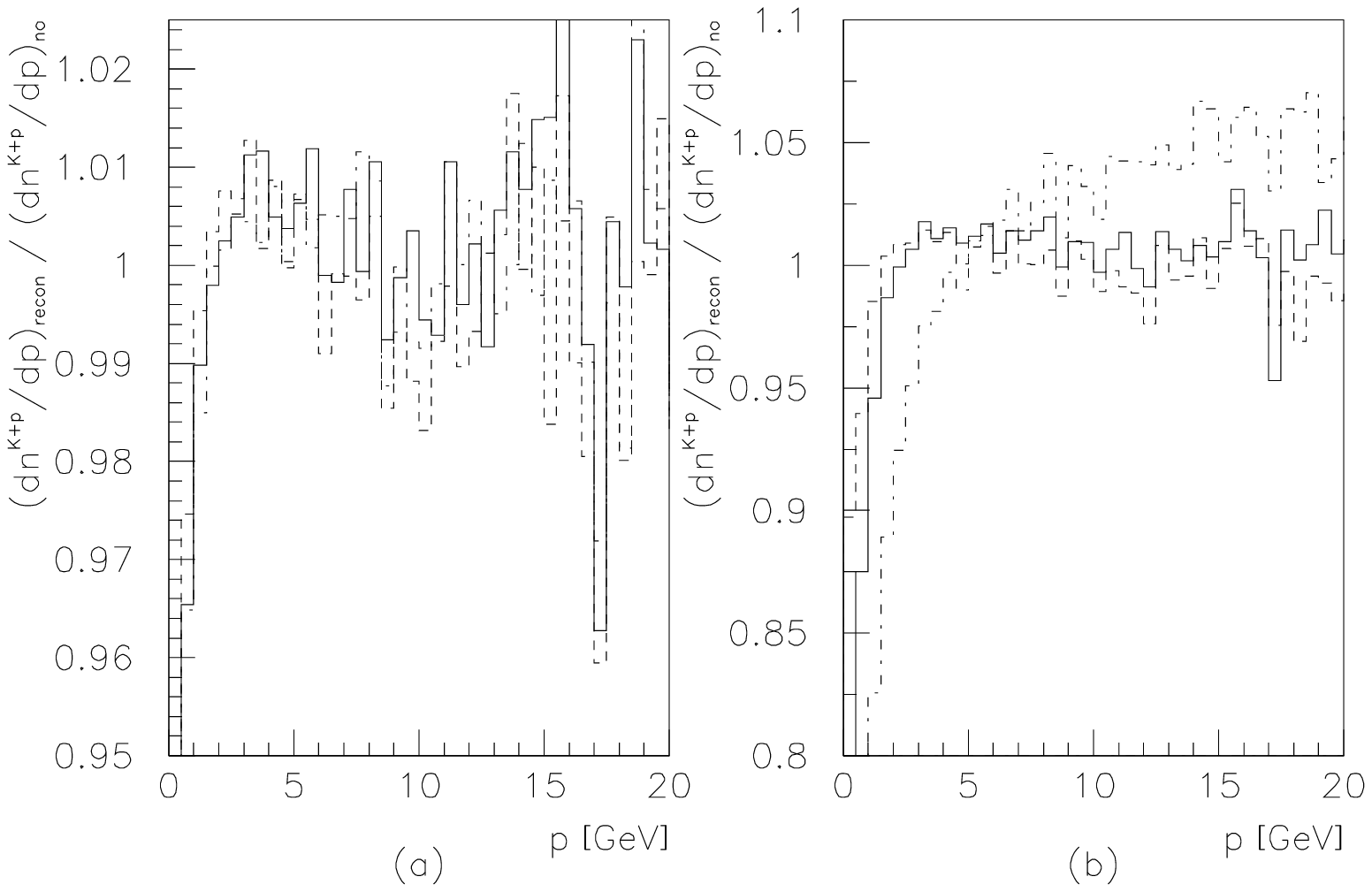}}
\end{center}
\vspace{-10mm}
\caption[dummy]{Ratio of reconnection to no-reconnection momentum
spectra $\d n^{\mathrm{K+p}} / \d p$ for heavy charged particles
($\K^\pm$ and $\p,\pbar$) only, with $p < 20$~GeV. Energy is 172~GeV; 
no ISR. Model curves as in Fig.~\protect\ref{fig1}.}
\label{fig3}
\begin{center}
%change for version 2: corrected curve for intermediate scenario
\mbox{\epsfig{file=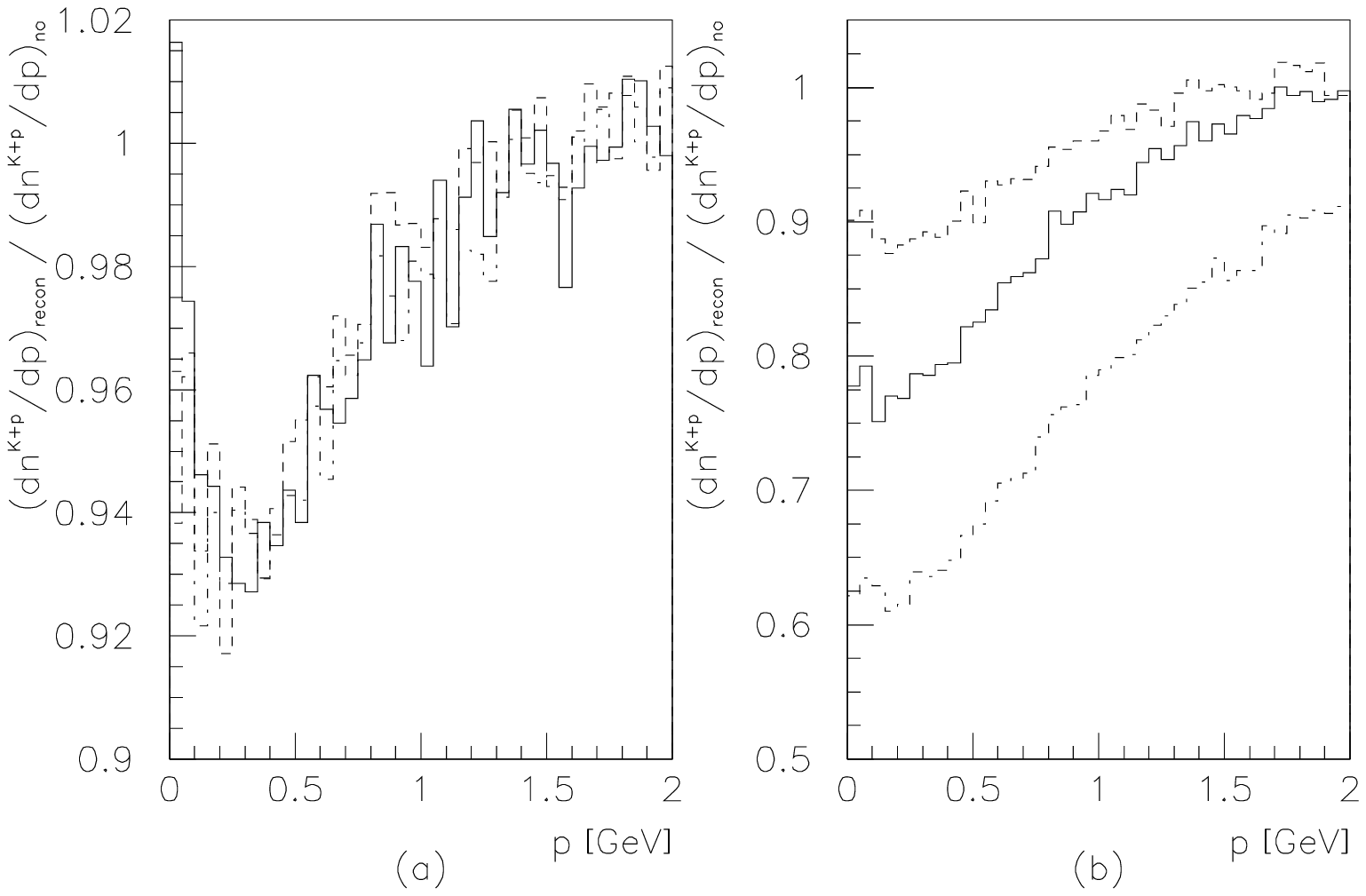}}
\end{center}
\vspace{-10mm}
\caption[dummy]{Ratio of heavy charged particle momentum spectra as in 
Fig.~\protect\ref{fig3}, for the low-momentum region $p < 2$~GeV.}
\label{fig4}
\end{figure}

To examine the r\^ole of the hadron mass we plot in Figs.~\ref{fig3} 
and \ref{fig4} the ratio of the $\K^{\pm}$ plus $\p,\pbar$ spectra for the 
same six models as before. This non-separated sample of heavy charged 
particles is experimentally more convenient to study than either of 
$\K$ or $\p$ separately, and carries the same physics message. 
As anticipated, the difference relative to the 
no-reconnection scenario becomes more marked in the heavy-charged-particle 
case. Moreover, the depletion now occurs in a somewhat wider momentum region. 
However, at the present stage we do not want to make too 
optimistic claims here. The chances of success of such studies
may well be endangered by low statistics (recall that $\K$+$\p$ are less
than 10\% of the soft particles) and the detection efficiency. 
All such issues need a lot of further work. For instance, a 
promising experimental approach is now suggested \cite{r26}, based on 
the combination of the $\d E / \d x$ and RICH
techniques for tagging low-momentum heavy hadrons.
In any case, one can cautiously say that if/when the reconnection 
signal is established safely in the charged-particle spectra, the
$\K$+$\p$ sample would provide additional instructive information.

\begin{figure}[t]
\begin{center}
%change for version 2: corrected curve for intermediate scenario
\mbox{\epsfig{file=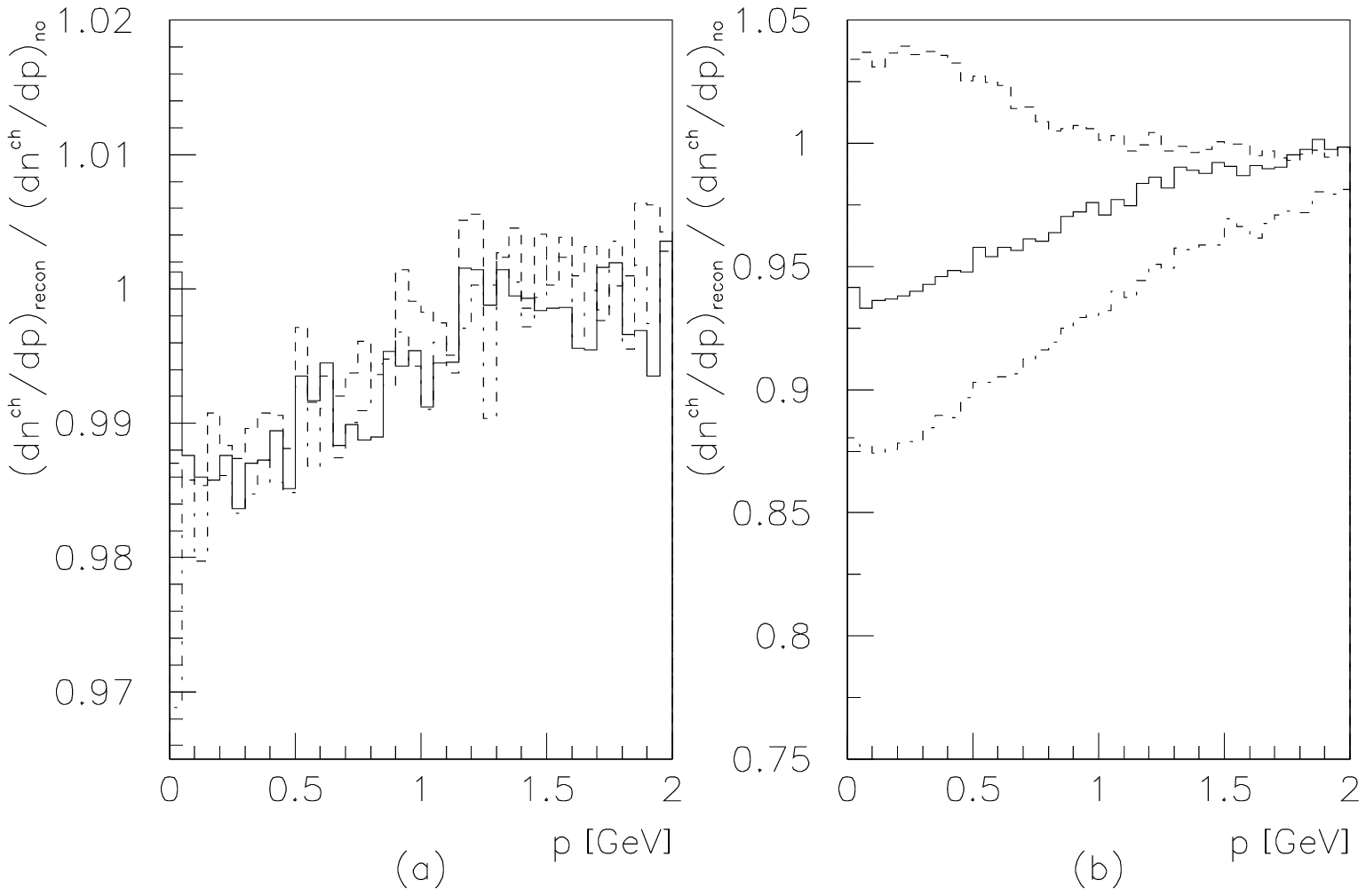}}
\end{center}
\vspace{-10mm}
\caption[dummy]{Ratio of charged momentum spectra as in 
Fig.~\protect\ref{fig2}, for the low-momentum region $p < 2$~GeV,
but at an energy of 195~GeV. No ISR.}
\label{fig5}
\end{figure}

As has been mentioned already, we expect a gradual decrease of the
cross-talk signal above threshold. In order to quantify this
effect we present in Fig.~\ref{fig5} the ratio of momentum spectra 
in the range $p < 2$~GeV at $\sqrt{s} = 195$~GeV. The numbers 
corresponding to this energy are shown in Tab.~\ref{tab1}.
The message from comparisons between the results at the two energies 
is quite clear: at $\sqrt{s} = 195$~GeV the differences drop down to 
about a half of what they were at 172~GeV. Recalling that the LEP2 
statistics leans towards higher energies, this result  naturally does 
not make the prospects of connectometry at LEP2 more optimistic.

For the $\W$ mass shift itself, the story is more complicated. At a
higher energy the two $\W$'s are more boosted apart, so a fixed change 
of low-momentum particles could have a larger impact on the $\W$ mass.
Over the LEP2 energy range the net effect, of a decreasing change of 
the momentum spectra with an increasing specific impact, is an 
essentially unchanged uncertainty in the $\W$ mass. That is, the shift
in individual models may move in either direction as a function of energy, 
but for the `envelope' of reasonable models there is no reason to expect 
neither a reduction nor an increase in the level of uncertainty. 

We show below some results of further studies, in order
to provide a more quantitative understanding  of  how large
effects are to be expected  under different requirements.

\begin{table}[t]
\begin{center}
\begin{tabular}{|l|c|c|c|c|r|}
\hline 
  & \multicolumn{4}{c|}{particle species and $p$ range (GeV)} &  \\
\hline
model & charged & charged & K$^{\pm}$ & $\p$,$\pbar$ &
$\langle \delta \MW \rangle$ \\ 
  & all & 0.15--0.7 & 0.15--0.7 & 0.15--0.7 & (MeV) \\ 
\hline
no-reconnection      &38.264 &  10.099 &   0.495 &   0.159 & ---  \\ \hline
I/no-rec             & 0.992 &   0.975 &   0.951 &   0.923 &    7 \\
II/no-rec            & 0.995 &   0.982 &   0.958 &   0.928 & $-5$ \\
II'/no-rec           & 0.992 &   0.976 &   0.952 &   0.922 & $-4$ \\ \hline
`GH'/no-rec          & 0.970 &   0.911 &   0.842 &   0.746 & $-55$ \\
%change for version 2: corrected numbers for intermediate scenario
%intermediate/no-rec  & 0.922 &   0.820 &   0.730 &   0.647 &  333 \\
Intermediate/no-rec  & 0.989 &   0.962 &   0.926 &   0.862 &   97 \\
%change for version 2: capitalize (minor typo for consistency)
%instantaneous/no-rec & 0.913 &   0.797 &   0.697 &   0.582 & 1054 \\
Instantaneous/no-rec & 0.913 &   0.797 &   0.697 &   0.582 & 1054 \\
\hline
\end{tabular}
\end{center}
\caption%
{Results on $\langle N_{\mathrm{ch}} \rangle$, 
$\langle N_{\mathrm{K}} \rangle$ and $\langle N_{\p,\pbar} \rangle$
in different momentum intervals. The reconnection numbers are normalized
to the no-reconnection numbers in the first row. Also the 
reconnection-induced $\W$ mass shifts $\langle \delta \MW \rangle$ are
shown. Energy is 172~GeV; no ISR.
\label{tab2}}
\begin{center}
\begin{tabular}{|l|c|c|c|c|r|}
\hline 
  & \multicolumn{4}{c|}{particle species and $p$ range (GeV)} &  \\
\hline
model & charged & charged & K$^{\pm}$ & $\p$,$\pbar$ &
$\langle \delta \MW \rangle$ \\ 
  & all & 0.15--0.7 & 0.15--0.7 & 0.15--0.7 & (MeV) \\ 
\hline
no-reconnection         & 38.385 &   9.357 &   0.430 &   0.124 &   ---  \\ \hline
I/no-rec                &  0.996 &   0.989 &   0.976 &   0.966 &     19 \\
II/no-rec               &  0.997 &   0.991 &   0.980 &   0.966 &  $-28$ \\
II$'$/no-rec            &  0.996 &   0.987 &   0.978 &   0.952 &  $-26$ \\ \hline
`GH'/no-rec             &  0.982 &   0.948 &   0.912 &   0.895 &     92 \\
%change for version 2: corrected numbers for intermediate scenario
%Intermediate/no-rec     &  0.927 &   0.858 &   0.825 &   0.838 &    307 \\
Intermediate/no-rec     &  1.008 &   1.030 &   1.064 &   1.147 &    316 \\
Instantaneous/no-rec    &  0.958 &   0.892 &   0.840 &   0.811 &   1473 \\ 
\hline
\end{tabular}
\end{center}
\caption%
{Results as in Tab.~\protect\ref{tab2}, but at 195~GeV.
No ISR.
\label{tab3}}
\begin{center}
\begin{tabular}{|l|c|c|c|c|r|}
\hline 
  & \multicolumn{4}{c|}{particle species and $p$ range (GeV)} &  \\
\hline
model & charged & charged & K$^{\pm}$ & $\p$,$\pbar$ &
$\langle \delta \MW \rangle$ \\ 
  & all & 0.15--0.7 & 0.15--0.7 & 0.15--0.7 & (MeV) \\ 
\hline
no-reconnection         & 38.388 &   9.445 &   0.436 &   0.127 &   --- \\ \hline
I/no-rec                &  0.995 &   0.986 &   0.974 &   0.969 &   36  \\
II/no-rec               &  0.996 &   0.987 &   0.977 &   0.953 & $-19$ \\
II$'$/no-rec            &  0.995 &   0.984 &   0.970 &   0.957 & $-21$ \\ \hline
`GH'/no-rec             &  0.980 &   0.943 &   0.905 &   0.878 &    80 \\
%change for version 2: corrected numbers for intermediate scenario
%Intermediate/no-rec     &  0.926 &   0.851 &   0.817 &   0.824 &   291 \\
Intermediate/no-rec     &  1.005 &   1.020 &   1.052 &   1.102 &   279 \\
Instantaneous/no-rec    &  0.953 &   0.880 &   0.825 &   0.779 &  1360 \\ 
\hline
\end{tabular}
\end{center}
\caption%
{Results as in Tabs.~\protect\ref{tab2} and \protect\ref{tab3}, but
at 195 GeV and with ISR included.
\label{tab4}}
\end{table}

Tab.~\ref{tab2} shows the numbers at $\sqrt{s} = 172$~GeV, without 
inclusion of ISR effects. 
The first line represents the reference numbers without
reconnection. In addition to the total multiplicity, we give the multiplicity 
in the range $0.15 < p < 0.7$~GeV, which roughly maximizes the size of
reconnection effects. (A smaller range would reduce statistics, a larger
reduce the average level of effects.)
The mass shift $\langle \delta \MW \rangle$ is defined as
\begin{equation}
\langle \delta \MW \rangle = \langle \MW \rangle_{\mathrm{recon}} -
  \langle \MW \rangle_{\mathrm{no-recon}}
\end{equation}
For illustration purposes we have chosen mass reconstruction method 3
of \cite{r6}.The basic idea of this method is to minimize the sum of 
deviations from an 80~GeV nominal mass,
$|\MW^{(1)}-80~\mathrm{GeV}| +|\MW^{(2)}-80~\mathrm{GeV}|$,
where  $\MW^{(1,2)}$ are the two reconstructed jet-jet masses. 
(The exact value used for the nominal mass is not critical; the idea is 
only to reject combinations that are clearly unreasonable.) The 
minimization is performed over the three distinct ways to pair the four 
jets. Note that the numbers in the last column in Tab.~\ref{tab2} do 
not coincide with the numbers in \cite{r6}: in the current study all 
$\W\W$ events are considered, while jet angular separation and 
minimum energy cuts were applied in \cite{r6}.

Apart from the spread of $\langle \delta \MW \rangle$ values among the 
four realistic models, also note that the $\langle \delta \MW \rangle$
%change for version 2: modified examples of the spread
%of the intermediate and instantaneous toy models differ by about a factor 
%of three,% 
of the intermediate toy model differs significantly from these,
although the depletion of low-momentum particles is fairly 
similar at 172 GeV. The instantaneous scenario gives the largest effects,
also e.g. if scaled down by a factor of 2 to bring momentum spectra
more in line with `GH' spectra.%
\footnote{Naively, the instantaneous mass shift could have been even 
larger. However, most events do not have a large amount of wide-angle 
radiation, where the coherence pattern of the radiation would be of 
importance. Also, most of the time the four quarks are not 
moving in directions close to each 
other and, when they are, the charge- and parity-symmetry-breaking structure 
of the $\W$ decays preferentially puts two quarks in the same hemisphere 
rather than a $\q$ from one $\W$ and a $\qbar$ from the other. Thus the 
reconnected systems are closer to the original ones than pure random.}
%change for version 2: part of sentence moved to above footnote.
%although the depletion of low-momentum particles is fairly 
%similar. This drives home the point that a depletion measurement does
This drives home the point that a depletion measurement does
not necessarily provide a correction procedure for the $\W$ mass 
determination. At best, an observation of no depletion could help reduce 
the systematic uncertainty that should be  assigned to the mass.

The message from comparison between the numbers in the `charged', 
`$\K^{\pm}$' and `$\p,\pbar$' columns in Tab.~\ref{tab2} is that there is 
no special need for a complete $\pi$/K/p separation --- one can rest 
content with a more relaxed separation $\pi$/non-$\pi$. (Primary produced
electrons and muons are of negligible importance, so we do not here 
address how they should be classified experimentally.)
We also remind that K$^0_{\mathrm{S}}$
behaves the same way as K$^{\pm}$ and that $\Lambda$ and other
strange baryons very closely follow the same pattern as $\p$.  
Therefore a rather inclusive tagging by secondary vertices (in the range
above charm and bottom decays) fills a similar function as a
non-$\pi$ sample of charged particles.
 
Since LEP2 statistics is biased towards higher energies, in 
Tab.~\ref{tab3} we give the numbers for $\sqrt{s} = 195$~GeV, 
still without ISR corrections. Comparing the ratios in Tab.~\ref{tab3} 
with those in  Tab.~\ref{tab2} one observes once more the anticipated 
reduction of the signal with center-of-mass energy.

One may worry that the presence of ISR might smear out momentum 
distributions, and thus dilute the reconnection signal. The
momentum range where the ratio of the spectra drops below unity might also
extend outwards. However, given that the boosts induced by the ISR are of 
rather moderate magnitude, at least on the scale considered here, the effects
are not expected to be large. To quantify the possible impact of ISR corrections, 
we give in  Tab.~\ref{tab4} the results corresponding to the same parameters as 
in Tab.~\ref{tab3}, but with ISR effects included. A comparison of the 
results in the two tables do not show any essential changes  in particle 
number ratios.  If anything, the ISR enhances the signals, by shifting the 
$\W\W$ system to a lower invariant mass, thus counteracting the general drop 
of effects with increasing energy. We note that the impact of ISR is small also 
at 172~GeV (not shown), as a trivial consequence of the reduced phase space for 
photon emission. So ISR corrections cannot make a significant impact on the 
conclusions of this paper. By contrast, the reconnection mass shift {\em is}
affected by ISR, both at 172 and 195~GeV, thus reflecting the non-negligible 
dependence of $\langle \delta \MW \rangle$ on the details of the (simulated) 
reconstruction procedures.

\begin{table}[t]
\begin{center}
\begin{tabular}{|l|c|c|c|c|c|r|}
\hline 
  & & \multicolumn{4}{c|}{particle species and $p$ range (GeV)} &  \\
\hline
model & survival &charged & charged & K$^{\pm}$ & $\p$,$\pbar$ &
$\langle \delta \MW \rangle$ \\ 
 & rate  & all & 0.15--0.7 & 0.15--0.7 & 0.15--0.7 & (MeV) \\ 
\hline
no-reconnection         & 0.343 &  37.207 &   9.683 &   0.479 &   0.157 &      --- \\ \hline
I/no-rec                & 0.344 &   0.992 &   0.978 &   0.966 &   0.935 &   $-1$ \\
II/no-rec               & 0.342 &   0.996 &   0.984 &   0.963 &   0.947 &   $-6$ \\
II$'$/no-rec            & 0.344 &   0.993 &   0.980 &   0.960 &   0.940 &  $-20$ \\ \hline
`GH'/no-rec             & 0.346 &   0.972 &   0.921 &   0.862 &   0.800 &  $-58$ \\
%change for version 2: corrected numbers for intermediate scenario
%Intermediate/no-rec     & 0.362 &   0.937 &   0.857 &   0.791 &   0.756 &    310 \\
Intermediate/no-rec     & 0.345 &   0.992 &   0.971 &   0.942 &   0.906 &     86 \\
Instantaneous/no-rec    & 0.355 &   0.929 &   0.831 &   0.750 &   0.676 &    607 \\ 
\hline
\end{tabular}
\end{center}
\caption%
{Results as in Tab.~\protect\ref{tab2}, but only events surviving
a thrust cut $0.8 < T < 0.9$. Energy is 172 GeV; no ISR.
\label{tab5}}
\begin{center}
\begin{tabular}{|l|c|c|c|c|}
\hline 
  & \multicolumn{4}{c|}{particle species and $p$ range (GeV)}  \\
\hline
model & charged & charged & K$^{\pm}$ & $\p$,$\pbar$ \\ 
  & all & 0.15--0.7 & 0.15--0.7 & 0.15--0.7 \\ 
\hline
no-reconnection         & 10.715 &   5.843 &   0.319 &   0.108 \\ \hline
I/no-rec                &  0.979 &   0.970 &   0.948 &   0.926 \\
II/no-rec               &  0.986 &   0.977 &   0.953 &   0.928 \\
II$'$/no-rec            &  0.982 &   0.972 &   0.950 &   0.927 \\ \hline
`GH'/no-rec             &  0.924 &   0.892 &   0.834 &   0.747 \\
%change for version 2: corrected numbers for intermediate scenario
%Intermediate/no-rec     &  0.791 &   0.779 &   0.709 &   0.644 \\
Intermediate/no-rec     &  0.963 &   0.950 &   0.918 &   0.861 \\
Instantaneous/no-rec    &  0.770 &   0.752 &   0.676 &   0.578 \\ 
\hline
\end{tabular}
\end{center}
\caption%
{The same observables as in Tab.~\protect\ref{tab2}, but only
for soft particles at least $30^{\circ}$ away from a jet direction.
Energy is 172~GeV; no ISR. 
\label{tab6}}
\end{table}

The main lesson from the analysis of fully inclusive spectra is not too
encouraging: the effects predicted within the realistic models in a 
real-life experiment seems to be quite small.
In order to enhance the signature for hadronic cross-talk one may wish to
examine more specific measures, i.e. search for the signal in
the regions of phase-space where the effects are expected to be most
pronounced.

It has been understood from the very beginning \cite{r5,r6,r7} 
that cross-talk should be enhanced in event configurations 
with large thrust $T$ (or, equivalently, small separation angle between 
initial cross-talking $\q\qbar$ pairs). To quantify the possible impact 
of the large-thrust selection on the low-momentum spectra we show in 
Tab.~\ref{tab5} the numbers for a trust cut of $0.8 < T < 0.9$ at 
172~GeV, without ISR effects included. The region $T > 0.9$ contains few 
$\W^+\W^-$ events and a large background from ordinary 
$\e^+\e^- \to \gamma^*/\Z^{0*} \to \q\qbar$ events, so is of no 
practical interest. The first column shows the number of $\W\W$ 
events surviving this cut. The ratios have little of the hoped-for 
enhancement; in some models the effects are even reduced. A more detailed
study of different thrust bins offer little hope: in the realistic models,
%change for version 2: remove intermediate scenario
%it is difficult to find any pattern at all. In the intermediate and 
%instantaneous models, the reduced effects are intertwined with a change
it is difficult to find any pattern at all. In the
instantaneous model, the reduced effects are intertwined with a change
of the thrust distribution itself. We remind that low thrust is not only 
a signal of well-separated $\W$ decay axes but also of energetic gluon 
emission or fluctuations in the fragmentation process towards higher
multiplicities. Therefore it is possible for a subdivision by thrust to
group events of similar multiplicity, and hence to reduce visible effects 
in the thrust-binned momentum spectra.

One could try to enhance the signal by measuring soft particles in
the interjet valleys only. Analogously to the string effect in three-jet
events of $\e^+\e^-$ annihilation \cite{r25} these regions of the 
phase space are the ones most affected by the Lorentz-boost effects.
Tab.~\ref{tab6} gives the results, corresponding to the additional 
selection requirement of particles lying at least $30^{\circ}$ away
from a jet direction, at 172~GeV.
Comparison between the numbers in Tabs.~\ref{tab2} and ~\ref{tab6} shows 
a certain rise of the signal in the latter case, but it is not dramatic.
To be more precise, the signal is increased about equally much,
relative to the results for all charged particles, by either of the
restrictions $0.15 < p < 0.7$~GeV and $\theta > 30^{\circ}$.
Combining the two restrictions does not give much further, presumably
since higher-momentum particles in the regions between the main jets
carry information about the perturbative gluon emission more than
about the soft cross-talk.

In summary, neither of the strategies considered above leads to a 
drastic improvement of the observability of the cross-talk signal. 
Finding better selection procedures  is still a challenging task, 
but at the moment  we  have no good proposal.

As a final exercise, we discuss below the possibility of using
$\Z^0$ decays for calibration of the low-momentum
spectra in the no-reconnection scenario.
Recall that, so far, the experimental strategy of searching for 
interconnection has been based on a comparison between the charged 
particle distributions in the (4q) decay channel
with those in twice the (2q) mode, e.g. \cite{r20,r21}. However, such a 
procedure has certain caveats. Firstly, the (2q) results have their 
own statistical and systematic uncertainties. Given that both event
samples are of comparable size, the statistical error in a comparison
of the (4q) with the (2q) channel is larger by at least a factor
$\sqrt{2}$ than the error in the (4q) channel itself. Secondly, the 
event selection cuts could introduce a bias between the 
purely hadronic and the mixed $\W\W$ decay modes.
Therefore, it looks quite appealing to try to develop a simple
model-independent procedure of replacing the measured soft-particle 
spectra in the individual, isolated $\W$ decay by spectra derived from 
the $\Z^0$ hadronic results. 

In order to acquire an appropriate `$\Z\Z$' reference set for comparisons 
with the $\W\W$ events, we have applied the following procedure.
First a $\W^+\W^-$ event is generated, using standard matrix elements 
as relevant at LEP2. Both $\W$'s decay hadronically. All such events 
are used in the analysis, i.e. there are no cuts to single out 
events with four well-separated jets. This is an attempt to stay in the 
spirit of a totally inclusive measurement, although experimentalists would 
certainly introduce some cuts to reduce background. The boost factors of 
the two W's and the directions of the  respective $\q\qbar$ event axis 
in the rest frame of its $\W$ is then found. These numbers
(two boost magnitudes and two polar angles, plus two azimuthal angles
that are irrelevant for inclusive $\d n / \d p$ spectra) 
are then applied to two $\Z^0$'s, each of which is taken to have the 
fixed mass $M_{\mathrm{Z}} = 91.2$~GeV.% 
\footnote{No attempt is made to rescale the momenta in $\Z^0$ events
to reduce the total energy from the $\Z^0$ to the $\W$ mass. Such a
rescaling could be relevant for a comparison of high-momentum 
particles, but would be misleading in the current studies. A rescaling
would imply, e.g., that the fragmentation $p_{\perp}$ would be
smaller in the rescaled $\Z^0$ events than in proper $\W$ events,
contrary to the universality normally assumed.}  
This means that, for each 
$\W^+\W^-$ event, a matching $\Z^0\Z^0$ event is found that can be 
analyzed in the same way. The $\Z^0$ flavours are chosen according to 
the relevant mixture, including b quarks. When ISR is included, each 
`$\Z\Z$' event is assumed to have the same set of ISR photons as the 
related $\W\W$ event from which the boosts and decay angles were adopted.

In an experimental analysis, the strategy above could be used to 
generate a large `$\Z\Z$' reference sample, with known boosts and decay 
angles, and making use of a $\Z^0$ generator (such as the 
{\sc Pythia}/{\sc Jetset} one here \cite{r27}) that has been extensively 
tuned to existing $\Z^0$ data. 
An alternative is to reconstruct boosts and angles from the observed
(4q) events, and to each such event assign a number of matching pairs
of observed $\Z^0$ events. (Given the enormous $\Z^0$ statistics and the 
small $\W^+\W^-$ one, there is no reason to stay with a one-to-one 
assignment of events.) 

\begin{figure}[p]
\begin{center}
\mbox{\epsfig{file=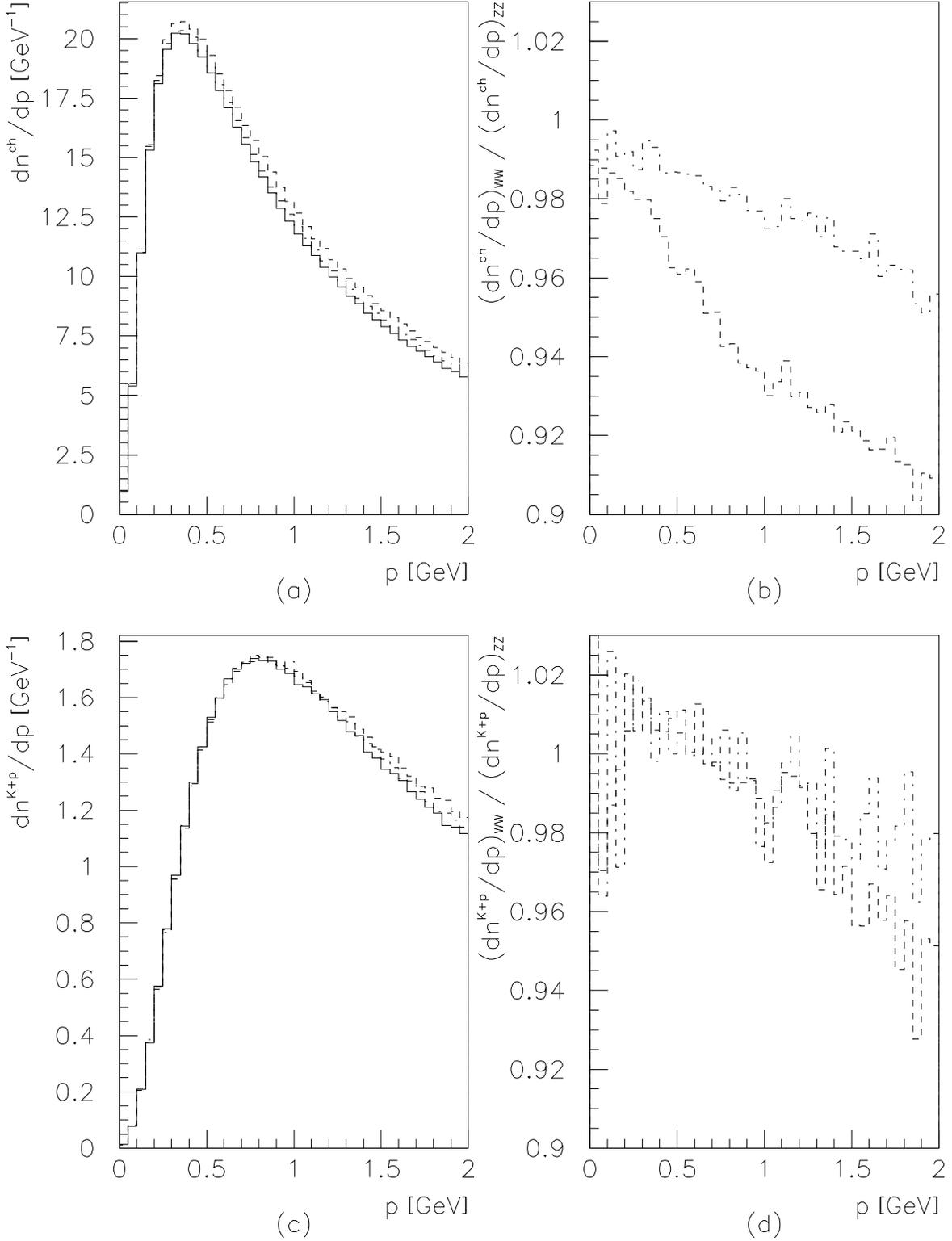}}
\end{center}
\vspace{-10mm}
\caption[dummy]{A comparison of momentum spectra in $\W\W$, `$\Z\Z$' 
and `$\Z\Z_{\mathrm{nb}}$' events. 
\textit{a)} $\d n^{\mathrm{ch}} / \d p$ spectra for the three models:
$\W\W$ full, `$\Z\Z$' dashed and `$\Z\Z_{\mathrm{nb}}$' dashed-dotted. 
\textit{b)} Ratios of these spectra: $\W\W$/`$\Z\Z$' dashed and
$\W\W$/`$\Z\Z_{\mathrm{nb}}$' dashed-dotted.  
\textit{c)} $\d n^{\mathrm{K}+\mathrm{p}} / \d p$ spectra for the three 
models: notation as in \textit{a)}.
\textit{d)} Ratios of these spectra: notation as in \textit{c)}.
All results are for 172~GeV; no ISR.}
\label{fig6}
\end{figure}

Some results from the application of the proposed recipe for 
comparing the $\W\W$ (no-reconnection scenario) events with 
the `$\Z\Z$' ones are given below.
The plots in Fig.~\ref{fig6} are for an 172~GeV energy, without ISR. 
Fig.~\ref{fig6}a,c show $\d n^{\mathrm{ch,K+p}} / \d p$ distributions 
in the range $0 < p < 2$~GeV for $\W\W$ and for `$\Z\Z$'. In 
Fig.~\ref{fig6}b,d the ratio of the two, $\W\W$/`$\Z\Z$', is displayed. 
The plots demonstrate that there is slightly more soft-particle 
multiplicity in the `$\Z\Z$' than in the  $\W\W$ events in the
$p < 1$~GeV region (an excess of $\sim 4$\% for all charged particles, 
but no net effect for kaons and protons). This is 
caused by the larger QCD evolution scale ($\Z^0$ mass as compared to the
$\W$ mass) and the presence of the $\b$-quark contribution in the 
former case. An examination of the full momentum range shows that
the $\W\W$/`$\Z\Z$' ratio always remains below unity. 

In order to understand the r\^ole of the primary quark flavour 
composition, we also show in Fig.~\ref{fig6} results where $\Z^0$ 
decays to $\b\bbar$ are excluded (`$\Z\Z_{\mathrm{nb}}$' events).
The difference is quite marked. For charged
particles with $p < 1$~GeV the $\W\W$/`$\Z\Z_{\mathrm{nb}}$' ratio is 
only $\sim 1$\% below unity. Recall that such a weak dependence of 
low-momentum spectra on the parton cascading scale is what one should 
expect on the basis of colour coherence in QCD branching processes, 
see e.g. \cite{r23} and references therein. 
For the $\K$+$\p$ case the $\W\W$/`$\Z\Z_{\mathrm{nb}}$' ratio becomes 
rather flat in a wide momentum interval, $p < 10$~GeV. 
For $p < 1$~GeV this ratio is again almost unity.

\begin{figure}[p]
\begin{center}
\mbox{\epsfig{file=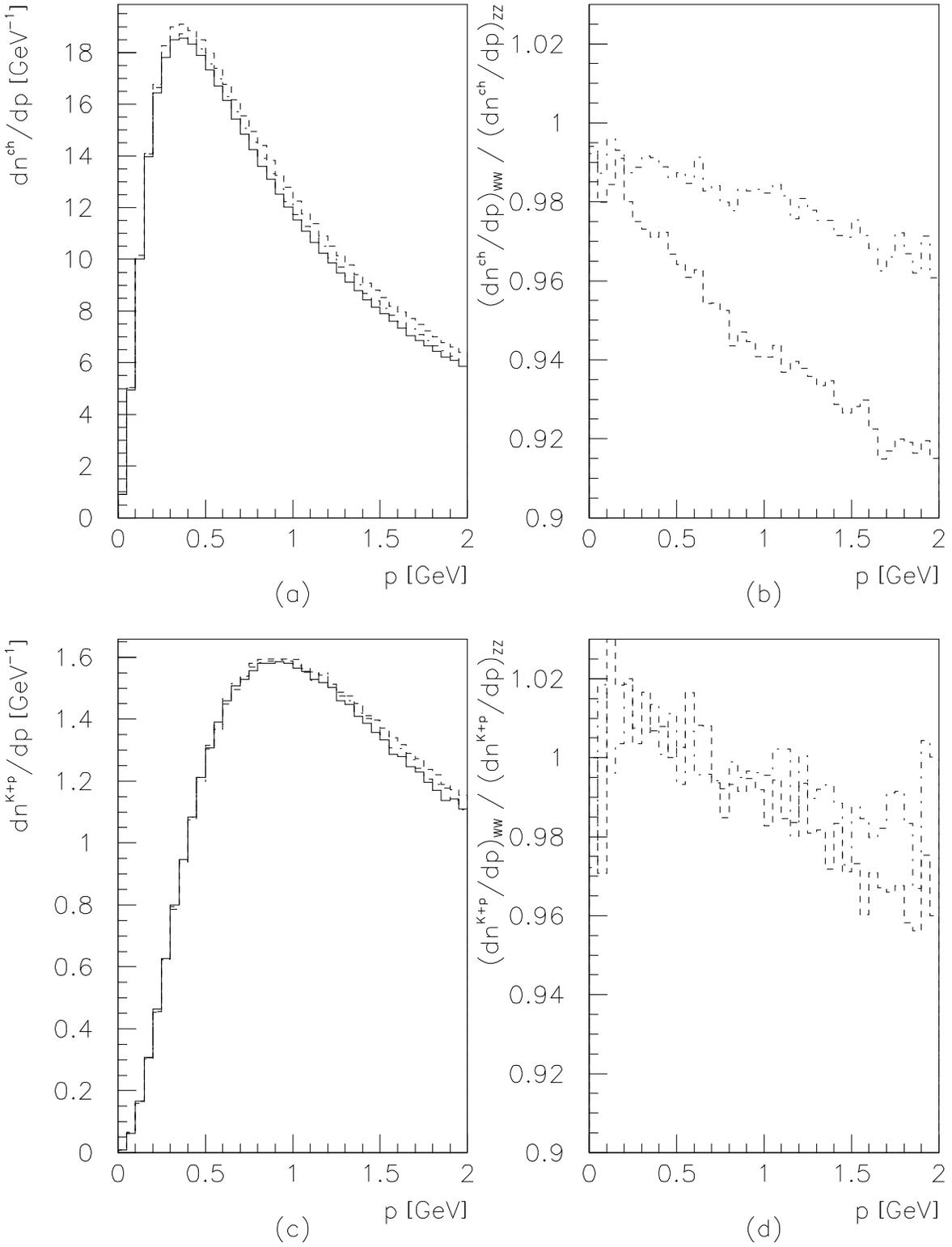}}
\end{center}
\vspace{-10mm}
\caption[dummy]{A comparison of momentum spectra in $\W\W$, `$\Z\Z$' 
and `$\Z\Z_{\mathrm{nb}}$' events, as in Fig.~\protect\ref{fig6}
but at an energy of 195~GeV. No ISR.}
\label{fig7}
\end{figure}

To test the possible energy dependence we plot in Fig.~\ref{fig7}
the same  observables as in Fig.~\ref{fig6}, but at 
$\sqrt{s} = 195$~GeV. A comparison of  these plots shows that,
at low momenta, the $\W\W$/`$\Z\Z$ ratio is almost energy-independent, 
but slightly  closer to unity for higher energies.
Studies in the energy range $\sqrt{s} = 160$--195~GeV 
demonstrate that, for $p < 1$~GeV, the ratio rises with $\sqrt{s}$ by at 
most 1\%.

To quantify further the applications of the proposed recipe for the 
calibration of soft-particle spectra we give in Tabs.~\ref{tab7} and
\ref{tab8} the numbers corresponding to different requirements,
at 172 and 195~GeV, respectively, with ISR included. In fact, as 
already noted above for the other studies, the inclusion or not of 
ISR has a negligible impact on any of the ratios between models.
Notice that the $\b$-quark contribution increases the multiplicity 
in the `$\Z\Z$' sample, but the  $\W\W$/`$\Z\Z$' ratio for soft 
kaons remains quite close to unity (within $\sim 1$\% accuracy) 
for both the `$\Z\Z$' and `$\Z\Z_{\mathrm{nb}}$' options. As for the
study of reconnection, thrust cuts do not have any significant impact on 
the $\W\W$/`$\Z\Z$' comparison.

\begin{table}[t]
\begin{center}
\begin{tabular}{|l|c|c|c|c|}
\hline 
  & \multicolumn{4}{c|}{particle species and $p$ range (GeV)}  \\
\hline
model & charged & charged & K$^{\pm}$ & $\p$,$\pbar$ \\ 
  & all & 0.15--0.7 & 0.15--0.7 & 0.15--0.7  \\ 
\hline
$\W\W$ (no-rec)               &  38.215 &  10.121 &   0.494 &   0.161  \\
`$\Z\Z$'                      &  41.449 &  10.450 &   0.496 &   0.157  \\
$\W\W$/`$\Z\Z$'               &   0.922 &   0.968 &   0.997 &   1.026  \\
`$\Z\Z_{\mathrm{nb}}$'        &  40.255 &  10.245 &   0.494 &   0.159  \\
$\W\W$/`$\Z\Z_{\mathrm{nb}}$' &   0.949 &   0.988 &   1.001 &   1.014  \\
\hline
\end{tabular}
\end{center}  
\caption%
{A comparison of soft-particle production between $\W\W$, `$\Z\Z$'
and `$\Z\Z_{\mathrm{nb}}$' events.
Energy is 172~GeV; with ISR. 
\label{tab7}}
\begin{center}
\begin{tabular}{|l|c|c|c|c|}
\hline 
  & \multicolumn{4}{c|}{particle species and $p$ range (GeV)}  \\
\hline
model & charged & charged & K$^{\pm}$ & $\p$,$\pbar$ \\ 
  & all & 0.15--0.7 & 0.15--0.7 & 0.15--0.7  \\ 
\hline
$\W\W$ (no-rec)               &  38.364 &   9.435 &   0.434 &   0.128 \\
`$\Z\Z$'                      &  41.433 &   9.721 &   0.437 &   0.124 \\
$\W\W$/`$\Z\Z$'               &   0.926 &   0.971 &   0.992 &   1.032 \\
`$\Z\Z_{\mathrm{nb}}$'        &  40.245 &   9.538 &   0.436 &   0.127 \\
$\W\W$/`$\Z\Z_{\mathrm{nb}}$' &   0.953 &   0.989 &   0.994 &   1.006 \\
\hline
\end{tabular}
\end{center}
\caption%
{Results as in Tab.~\protect\ref{tab7}, but at 195~GeV.
\label{tab8}}
\end{table}

The message from examining the results, presented in Figs.~\ref{fig6} 
and \ref{fig7} and in Tabs.~\ref{tab7} and \ref{tab8}, is that the 
proposed procedure for calibration of the $\W\W$ no-reconnection sample 
should allow a reference for the low-momentum spectra within an 
accuracy of $\sim 3$\% for charged particles and $\sim 1$\% for 
$\K$+$\p$. It is also worthwhile to mention that the  $\W\W$/`$\Z\Z$' 
ratio is almost always slightly below unity.

Bearing in mind that, within the realistic scenarios, the expected depletion
of low-momentum spectra as compared to the no-reconnection case is 
$\sim 2$\% for charged particles and $\sim 5$\% for $\K$+$\p$, the 
advocated calibration recipe, at first sight, does not look safe.
However, we would like to recall that the essential part of the
deviation of the  $\W\W$/`$\Z\Z$' from unity is caused by the 
$\b$-quark effect. The latter is quite well controllable within the 
existing Monte Carlo models, so it looks like a straightforward 
procedure to correct for it. Conceivably one would even dare trust 
generators to have (most of) the remaining differences between $\W\W$
and `$\Z\Z$' under control, so that further corrections could be
applied. The situation is even more favourable for 
$\K$+$\p$, where the reconnection-induced shifts are expected to be larger 
and the deviation of the ratio from unity is smaller.
So there could be some prospects of a reliable  application of such a 
normalization recipe, but further work (in particular, on the experimental 
side) is needed.

\section{Conclusion}

One of the main challenges facing experimental studies on the $\W$-mass
reconstruction at LEP2 concerns systematic uncertainties caused by the 
strong interactions between the $\W^+\W^-$ decay products. This subject has 
recently attracted much attention, see for reviews \cite{r1,r2,r3,r4,r20}.

The existing theoretical literature, based on quite different 
philosophies, shows a rather wide range of expectations for the shift 
in $\MW$, from a few MeV to several hundred MeV. The cross-talk 
between the W's may have an impact on various other properties of hadronic  
$\W^+\W^-$ events as well.
 
Different hypotheses about the confinement dynamics may  lead to 
different expectations for the final-state event characteristics.
So, in principle, the experimental tests of hadronic interconnection 
between the $\W$'s --- connectometry --- could provide a new laboratory
for probing the structure of the QCD vacuum. This important issue was 
first addressed in \cite{r5} and since then has become a subject of 
intensive discussions.
 
In order to establish the evidence for a cross-talk one has to find
an observable which, on the one hand, proves to be quite sensitive to 
this effect and, on the other hand, could allow rather straightforward
interpretation. The necessary requirements for such a connectometer are  
that the no-reconnection predictions should be  well understood, and 
that the expected signal is strong enough to be detectable within the 
limited statistics of LEP2. The latter is, by  no means, a simple task.

In the last year we have witnessed several experimental attempts to find
an evidence of interconnection, see \cite{r20,r21}. At the present level 
of statistics, no indication has been found.
   
This paper attempts to quantify the expectations based on the string
hadronization model \cite{r19} in terms of the distributions of 
low-momentum hadrons. This idea is motivated by an observation \cite{r6} 
that it is the soft particles that are  most sensitive to hadronic 
cross-talk. Essential advantages of such an approach to connectometry 
is that here the no-reconnection case can be well described, and that 
there is  no (direct) dependence on the jet reconstruction method or 
event selection strategy.

Studies are presented for six reconnection scenarios. Model I is based
on a bag-like model for strings, and II and II$'$ on a vortex-line
approach. In II$'$ and `GH' reconnections have to reduce the string 
length. For comparisons with the first three models, which have a
25--35\% reconnection rate, the `GH' deviations
from the no-reconnection reference should be scaled down by about a factor
of 3. An unscaled `GH' with (close to)
unit reconnection probability is to be considered as an extreme. This is
even more the case for the intermediate and instantaneous scenarios, which 
should be viewed as toy models, convenient for technical and reference
purposes only. Let us therefore be very clear here: the last three lines 
of Tabs.~\ref{tab1}--\ref{tab6} are so easy targets and of so small 
intrinsic value that no prize will be handed to experimentalists shooting 
them down. The real targets are represented by the I, II and II$'$ 
(and `GH'/3) numbers. 

It is of some interest to note that, for the four realistic models, the 
depletion of the low-momentum particle spectrum rather closely agree.
Since the depletion should scale proportionately to the reconnection
probability, we may here have a good gauge for this probability.  
But, as already emphasized, the models still disagree on the value
of the $\W$ mass shift, so we cannot offer a recipe for turning a
given level of depletion into a correction on the $\W$ mass.

The designated objective of LEP2 is to deliver 500~pb$^{-1}$ of
luminosity. This means about 4000 $\W\W \to 4\q$ events per experiment.
Noting that the number of charged particles in the 0.15--0.7~GeV momentum
bin follows a (close to) Poissonian distribution with a mean of  
$\approx 10$, this would give a statistical error of
$\delta n \approx \sqrt{10}/\sqrt{4000} = 0.05$ and a relative
error of $\delta n/n \approx 0.005$. The expected effect in realistic
models is $\sim 0.02$ at 172~GeV and $\sim 0.01$ at 195~GeV.
From a purely statistical point of view we could therefore expect at
most a $4\sigma$ deviation and more likely $2\sigma$, given the intended 
higher energy of future runs. It would be difficult to claim success based
on such numbers. There are two points in favour, however. One is the
presence of four experiments with independent statistical samples. 
The other is that we already believe to know the sign of the effect
we are looking for.

Of course, given the marginal statistical significance of a signal,
it is important to limit systematic errors. At first sight, this 
seems rather difficult. The obvious control sample of (2q) decays of
$\W$ pairs suffers from as severe a statistics problem as the (4q)
decays; actually a factor 2 worse since one needs two (2q) events
to simulate one (4q) one. It would therefore be an advantage if 
instead one could use the essentially unlimited $\Z^0$ sample as reference.
We have shown that a straight comparison is not accurate enough,
but that a correction for $\Z^0 \to \b\bbar$ decays almost brings one
there. Presumably generators are good enough at modelling differences
between $\W$ and $\Z$ decays that one would trust them also for some
further corrections, making the $\Z^0$ calibration method the best
bet and probably bringing this kind of systematic errors below the 
level of statistical ones. Other kinds of systematic errors are not 
considered in this article but may pose formidable challenges to the 
experimentalist. Of some small consolation is that our use of rather 
inclusive quantities implies a limited dependence on the details of the 
event selection cuts. 

As an alternative to considering all charged particles, one may focus
on the $\K$+$\p$ subsample in the 0.15--0.7~GeV momentum range.
The relative statistical error then is increased to
$\delta n/n \approx 0.02$, and the expected signal to $\sim 0.05$ at
172~GeV and $\sim 0.025$ at 195~GeV. From a statistical point of view
this is a clear degradation, but it appears that the calibration      
to the $\Z^0$ reference sample can be made with higher accuracy. While 
maybe not a signal in its own right, $\K$+$\p$ thus might appear in the 
`supporting cast', should a signal be found for all charged particles
in the low-momentum range. Another similar example is the number of
particles, at all momenta, that are not within $30^{\circ}$ of a jet,
while cuts on the thrust of events does not seem to offer anything.

Our choice of momentum range 0.15--0.7 GeV could be modified. The lower 
cut is purely a matter of convenience, since so few particles have 
$p < 0.15$~GeV anyway. The upper cut could be increased to 1~GeV, say, 
giving a decrease in the average level of the signal but an increase in 
the  number of particles contributing to it. The net effect is 
a slight increase in the purely statistical significance,
within the context of the reconnection models studied, but probably
at the expense of a somewhat worsened systematic uncertainty. A 
restriction to $p < 0.4$~GeV, on the other hand, reduces the statistics
so much that the increase in the average level of the signal could
not really compensate. Anyway, the message is that experimentalists 
should feel free to pick cuts based on a complete overview of all
issues, experimental and theoretical, while we here have only attempted 
to discuss the latter ones. Studies for the $\W$ mass determination,
in particular, may require rather special event selection cuts. More
relaxed cuts could then be considered for the inclusive spectrum, 
but is not an absolute prerequisite.

Our studies do not encourage a too optimistic prognosis concerning the 
prospects  of connectometry on the basis of low-momentum spectra, even 
having the whole aimed-for statistics of LEP2. The best we can hope is 
that the expected signal would be at the edge of observability. In such 
a case one would need a lot of hard work (and good luck) in order to 
detect the signal reliably. However, we would like to emphasize that, 
given the present lack of deep understanding of the non-perturbative 
QCD dynamics, it is only experiment that could lead the way and may 
cast light on the challenging issues of the hadronic cross-talk.

We would like to make it absolutely clear that the nonobservation of 
the reconnection effects on the low-momentum spectra, by no means, 
indicates their nonexistence. Most likely it may just mean that the 
``queen of observables'' is still to be nominated.

It may be useful here to recall some other ideas. 
We are not saying that either of them are offering larger hope than the 
currently studied one, but they could add pieces to the jigsaw puzzle:
\begin{Itemize}
\item In the string model, particles are produced at limited transverse 
momenta with respect to the string direction (in the rest frame of the
string). One may therefore attempt an event-by-event reconstruction of 
the colour topology by a minimization of the total transverse momentum
with respect to the potential sets of strings \cite{ENTS}. The inclusion 
of parton-shower activity smears the signal significantly, but still
effects are at a comparable level of statistical significance to the ones 
reported here. The chances of controlling systematics appear to be worse,
however. Also in this approach most of the information content is 
sitting in low-momentum particles, roughly out to 3~GeV. 
\item A more inclusive test of colour topology is the azimuthal 
distribution of particles around jet directions \cite{r6}.
Since the two jets of a $\W$ are approximately back-to-back, 
to zeroth approximation one would expect a flat distribution.
If a reconnection occurs, the new strings would be spanned
in the direction of the two jets from the other $\W$. 
In practice one could, for each jet, define $\varphi = 0$ 
as the direction to the nearest other jet. The signal for reconnection
would then be an enhanced particle production close to $\varphi = 0$ 
and $\varphi = \pi$, and a reduced one at around $\varphi = \pi/2$.
However, this effect has to be extracted from a much larger non-isotropy
induced by simple kinematical effects, e.g. that the jet clustering
procedure depletes the particle production in the direction
towards the next nearest jet. It has not been studied whether the signal
could be enhanced by suitable cuts, e.g. a restriction to low-momentum
particles.
\item A variant of the azimuthal anisotropy study is to concentrate on
activity out of the event plane of the two jets of a $\W$. In the absence
of reconnections, and neglecting hard perturbative gluons, particle
production is concentrated to the event plane of each $\W$. A 
reconnection would increase the activity out of the plane.
\item The reduced low-momentum particle production has to be compensated 
by an increase elsewhere, in order to conserve energy. We have above noted
that, in the realistic models, most of the effect occurs at intermediate 
momenta, say 1.5--5~GeV. At higher momenta both the original and the 
reconnected strings line up particles with the jet directions and have 
little memory of the central overlap region. The compensation effects
are rather small, since fewer particles need be added at intermediate 
momenta to give the same amount of energy as the one taken away at
low momenta. Given that experimentalists also measure in this range, 
there is a  possibility to increase the potential significance of a 
signal. The calibration to a no-reconnection scenario, e.g. based on
$\Z^0\Z^0$ events, would be even more delicate in this region, however.
\end{Itemize}

It is important to bear in mind, that even if/when the evidence for the
reconnection signal is established, it may  be quite a long way before
one would be able to draw a convincing conclusion concerning the 
$\W$-mass shift.
 
Finally, let us recall that the reconnection probability is a free 
parameter in the majority of the models. So if forthcoming results 
from LEP2 show shifts which are significantly greater than our 
expectations, this could signal that the reconnection probability 
is larger than in the models adopted in this paper.

\subsection*{Acknowledgments}

We would like to thank A.~De Angelis, M.~Battaglia, A.~Blondel, D.~Liko,
N.~Neufeld, R.~Orava, B.~Pietrzyk, A.~Tomaradze, N.~Watson and B.R.~Webber 
for useful discussions. VAK thanks the Theory Division of CERN for the 
warm hospitality during the course of this work. This work was supported 
in part by the EU Fourth Framework Programme `Training and Mobility of 
Researchers', Network `Quantum Chromodynamics and the Deep Structure of 
Elementary Particles', contract FMRX--CT98--0194 (DG 12 -- MIHT).
%change for version 2: add further acknowledgement
A special thank to D. van Dierendonck and P. de Jong for finding
a mistake in our original numbers for the intermediate scenario.

\end{document}